\documentclass[amsmath,amssymb,epsfig]{ws-procs9x6}

\begin{document}

\title{Anomaly Matching and Low Energy Theories at High Matter Density}

\author{Francesco Sannino}

\address{${\rm NORDITA}$ \\
Blegdamsvej 17, Copenhagen \O DK-2100, Denmark\\
E-mail: francesco.sannino@nbi.dk}

\maketitle

\abstracts{I provide new arguments supporting the validity of the
t'Hooft anomaly conditions at non zero quark chemical potential.
These constraints strengthen the quark-hadron continuity scenario.
Finally I review the 2SC effective Lagrangian for color
superconductivity.}

\section{Introduction}
To understand the low energy physics of strongly interacting
theories such as QCD and QCD like theories, where perturbation
theory is not applicable, effective Lagrangians constructed with
the aid of the global symmetries of the underlying theory play a
relevant role. To constrain the low energy dynamics at zero quark
chemical potential t'Hooft anomaly matching conditions are much
used.

While low energy effective Lagrangians at non zero quark chemical
potential for QCD like theories have been widely used in
literature, the consequences of the t'Hooft anomaly conditions in
this regime are still not fully explored. Here I provide new
arguments, not discussed in literature, supporting the validity of
the anomaly mathcing conditions at non zero chemical potential.
These constraints strengthen the quark-hadron continuity scenario.
Finally I briefly review the 2SC effective low energy theory.

\section{Color Superconductivity and the QCD Phase Diagram}
 At zero temperature but very high quark chemical potential
strong interactions favor the formation of quark-quark condensates
in the color antisymmetric channel \cite{REV}. Possible physical
applications are related to the physics of compact objects
\cite{REV}, supernovae explosions \cite{HHS} as well as to the
Gamma Ray Bursts puzzle\cite{OS}. Recently these ideas have been
investigated in detail in \cite{Blaschke:2003yn}.

\begin{figure}[htb]
\parbox[t]{5.8 cm} {\leavevmode
\begin{minipage}[c]{5.8cm}
\vspace*{-4.5cm} 
  \epsfxsize 5.8 cm \epsfysize 6 cm
  \epsfbox{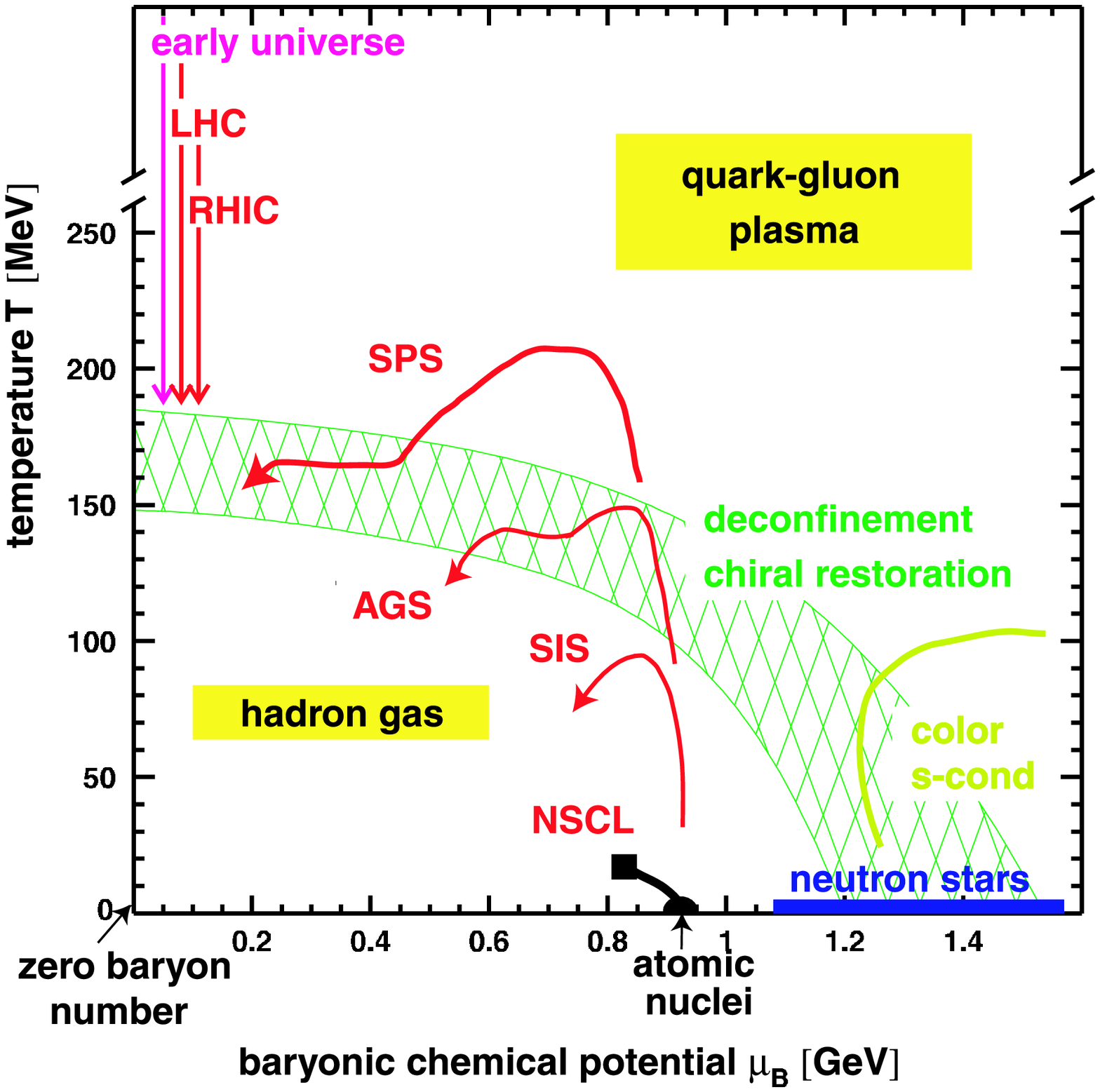}
\end{minipage}}
\ \hskip .2cm   \ \parbox[t]{5 cm} { \leavevmode \epsfxsize 5cm
\epsfysize 4 cm \epsfbox{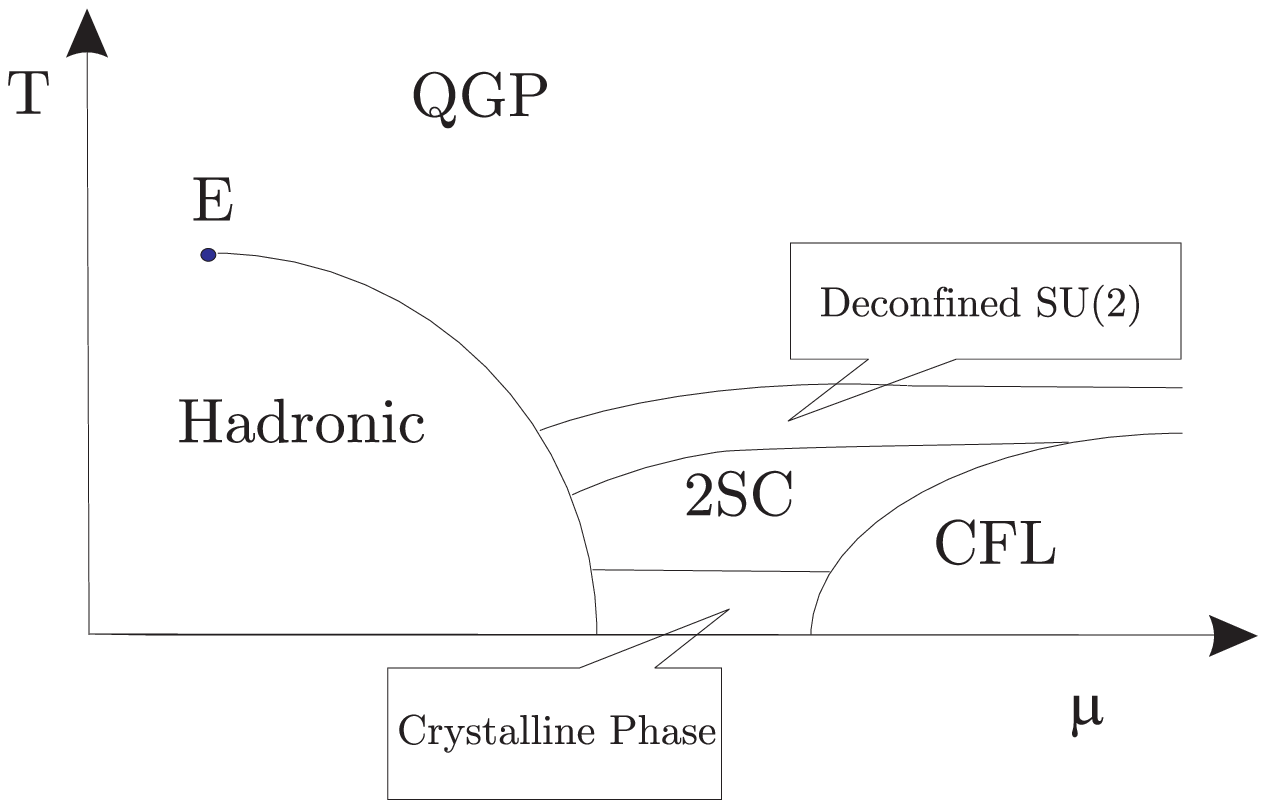} }\caption[] {Left Panel:Possible
physical applications and region of the QCD phase diagram explored
by different experiments. Right Panel: An oversimplified cartoon
of the theoretical QCD phase diagram. The crystalline phase may
exists only if we have different chemical potentials for the up
and down quarks. The deconfined $SU_c(2)$ corresponds to the phase
where we have gapped up and down quarks and the remaining unbroken
$SU_c(2)$ of color deconfined.}\label{tmu}
\end{figure}

According to the number of light flavors in play we have different
phases. A theoretical picture of the QCD phase diagram is
presented in the second panel of Fig.~\ref{tmu} while in the first
panel the relevant experiments and possible physical applications
are displayed \cite{Heinz:2001ax}.

\subsection{Color Flavor Locked Phase}
\label{3f}

{}For $N_{f}=3$ light flavors at very high chemical potential
dynamical computations suggest that the preferred phase is a
superconductive one and the following ansatz for a quark-quark
type of condensate is energetically favored:
\begin{equation}
\epsilon ^{\alpha \beta }<q_{L\alpha ;a,i}q_{L\beta ;b,j}>\sim
k_{1}\delta _{ai}\delta _{bj}+k_{2}\delta _{aj}\delta _{bi}\ .
\label{condensate}
\end{equation}
\noindent A similar expression holds for the right transforming
fields. The Greek indices represent spin, $a$ and $b$ denote color
while $i$ and $j$ indicate flavor. The condensate breaks the gauge
group completely while locking the left/right transformations with
color. The final global symmetry group is $SU_{c+L+R}(3)$, and the
low energy spectrum consists of $9$ Goldstone bosons.

The low energy effective theory for 3 flavors (CFL) has been
developed in \cite{CG}. We refer to \cite{REV,Casalbuoni} for a
complete summary and review of this phase.

\subsection{2 SC General Features}
\label{tre}

QCD with 2 massless flavors has gauge symmetry $SU_{c}(3)$ and
global symmetry
\begin{equation}
SU_{L}(2)\times SU_{R}(2)\times U_{V}(1)\ .
\end{equation}
At very high quark density the ordinary Goldstone phase is no
longer favored compared with a superconductive one associated to
the following type of diquark condensates:
\begin{equation}
\langle L{^{\dagger }}^{a}\rangle \sim \langle \epsilon
^{abc}\epsilon ^{ij}q_{Lb,i}^{\alpha }q_{Lc,j;\alpha }\rangle \
,\qquad \langle R{^{\dagger }}^{a}\rangle \sim -\langle \epsilon
^{abc}\epsilon ^{ij}q_{Rb,i;\dot{\alpha}
}q_{Rc,j}^{\dot{\alpha}}\rangle \ ,
\end{equation}
If parity is not broken spontaneously, we have $\left\langle
L_{a}\right\rangle =\left\langle R_{a}\right\rangle =f\delta
_{a}^{3}$, where we choose the condensate to be in the 3rd
direction of color. The order parameters are singlets under the
$SU_{L}(2)\times SU_{R}(2)$ flavor transformations while
possessing baryon charge $\frac{2}{3}$. The vev leaves invariant
the following symmetry group:
\begin{equation}
\left[ SU_{c}(2)\right] \times SU_{L}(2)\times SU_{R}(2)\times
\widetilde{U}_{V}(1)\ ,
\end{equation}
where $\left[ SU_{c}(2)\right] $ is the unbroken part of the gauge
group. The $\widetilde{U}_{V}(1)$ generator $\widetilde{B}$ is the
following linear combination of the previous $U_{V}(1)$ generator
$B$ and the broken diagonal generator of the $SU_{c}(3)$ gauge
group $T^{8}$:
$\displaystyle{\widetilde{B}=B-\frac{2\sqrt{3}}{3}T^{8}={\rm
diag}(0,0,1)}$. The quarks with color $1$ and $2$ are neutral
under $\widetilde{B}$ and consequently so is the condensate.

\section{Anomaly Matching Conditions}
The superconductive phase for $N_{f}=2$ possesses the same global
symmetry group as the confined Wigner-Weyl phase. The ungapped
fermions have the correct global charges to match the t' Hooft
anomaly conditions as shown in \cite{S}. Specifically the
${SU(2)_{L/R}}^2\times U(1)_V$ global anomaly is correctly
reproduced in this phase due to the presence of the ungapped
fermions. This is so since a quark in the 2SC case is surrounded
by a diquark medium (i.e. $q \left< q q\right>$) and behaves as a
baryon.
\begin{eqnarray} \nonumber
\left(%
\begin{array}{c}
  u \\
  d \\
\end{array}%
\right)_{color = 3} \quad \sim \quad \left(%
\begin{array}{c}
  p \\
  n \\
\end{array}%
\right)\ .
\end{eqnarray}
The validity of the t'Hooft anomaly conditions at high matter
density have been investigated in \cite{S,Hsu:2000by}. A delicate
part of the proof presented in \cite{Hsu:2000by} is linked
necessarily to the infrared behavior of the anomalous three point
function. In particular one has to show the emergence of a
singularity (i.e. a pole structure). This pole is then interpreted
as due to a goldstone boson when chiral symmetry is spontaneously
broken.

One might be worried that, since the chemical potential explicitly
breaks Lorentz invariance, the gapless (goldstone) pole may
disappear modifying the infrared structure of the three point
function. This is not possible. Thanks to the Nielsen and Chadha
theorem \cite{Nielsen}, not used in \cite{Hsu:2000by}, we know
that gapless excitations are always present when some symmetries
break spontaneously even in the absence of Lorentz
invariance\footnote{Under specific assumptions which are met when
Lorentz invariance is broken via the chemical potential.}. Since
the quark chemical potential is associated with the barionic
generator which commutes with all of the non abelian global
generators the number of goldstone bosons must be larger or equal
to the number of broken generators. Besides all of the goldstones
must have linear dispersion relations (i.e. are type I
\cite{Nielsen}). This fact not only guarantees the presence of
gapless excitations (justifying the analysis made in
\cite{Hsu:2000by} on the infrared behavior of the form factors)
but demonstrates that the pole structure due to the gapless
excitations needed to saturate the triangle anomaly is identical
to the zero quark chemical potential one in the infrared.

It is also interesting to note that the explicit dependence on the
quark chemical potential is communicated to the goldstone
excitations via the coefficients of the effective Lagrangian (see
\cite{Casalbuoni} for a review). {}For example $F_{\pi}$ is
proportional to $\mu$ in the high chemical potential limit and the
low energy effective theory is a good expansion in the number of
derivatives which allows to consistently incorporate in the theory
the Wess-Zumino-Witten term \cite{S} and its corrections.

The validity of the anomaly matching conditions have far reaching
consequences. Indeed, in the three flavor case, the conditions
require the goldstone phase to be present in the hadronic as well
as in the color superconductive phase supporting the quark-hadron
continuity scenario \cite{Schafer:1998ef}. At very high quark
chemical potential the effective field theory of low energy modes
(not to be confused with the goldstone excitations) has positive
Euclidean path integral measure \cite{Hong:2002nn}. In this limit
the CFL is also shown to be the preferred phase. Since the
fermionic theory has positive measure only at asymptotically high
densities one cannot use this fact to show that the CFL is the
preferred phase for moderate chemical potentials. This is possible
using the anomaly constraints.

While the anomaly matching conditions are still in force at non
zero quark chemical potential \cite{S} the {\it persistent mass}
condition \cite{Preskill:1981sr} ceases to be valid. Indeed a
phase transition, as function of the strange quark mass, between
the CFL and the 2SC phases occurs.

We recall that we can saturate the t'Hooft anomaly conditions
either with massless fermionic degrees of freedom or with gapless
bosonic excitations. However in absence of Lorentz covariance the
bosonic excitations are not restricted to be fluctuations related
to scalar condensates but may be associated, for example, to
vector condensates \cite{Sannino:2002wp}.

\section{2SC Effective Low Energy Theory}

The spectrum in the 2SC state is made of 5 massive Gluons with a
mass of the order of the gap, 3 massless Gluons confined (at zero
temperature) into light glueballs and gapless up and down quarks
in the direction (say) 3 of color. \subsection{The 5 massive
Gluons} The relevant coset space $G/H$ \cite{CDS,{CDS2001}} with
\begin{eqnarray}
G=SU_{c}(3)\times U_{V}(1)\ , \quad   {\rm and} \quad
H=SU_{c}(2)\times \widetilde{U}_{V}(1) \end{eqnarray} is
parameterized by:
\begin{eqnarray}
{V}=\exp (i\xi ^{i}X^{i})\ ,
\end{eqnarray}
where $\{X^{i}\}$ $i=1,\cdots ,5$ belong to the coset space $G/H$
and are taken to be $X^{i}=T^{i+3}$ for $i=1,\cdots ,4$ while
\begin{eqnarray}
X^{5}=B+\frac{\sqrt{3}}{3}T^{8}={\rm
diag}(\frac{1}{2},\frac{1}{2},0)\ . \label{broken}
\end{eqnarray}
$T^{a}$ are the standard generators of $SU(3)$. The coordinates
\begin{eqnarray} \nonumber
\xi ^{i}=\frac{\Pi ^{i}}{f}\quad i=1,2,3,4\ ,\qquad \xi
^{5}=\frac{\Pi ^{5}}{\widetilde{f}}\ ,
\end{eqnarray}
via $\Pi $ describe the Goldstone bosons which will be absorbed in
the longitudinal components of the gluons. The vevs $f$ and
$\widetilde{f}$ are, at asymptotically high densities,
proportional to $\mu $. ${V}$ transforms non linearly:
\begin{eqnarray}
{V}(\xi )\rightarrow u_{V}\,g \,{V}(\xi )\,h^{\dagger }(\xi
,g,u)\,h_{\widetilde{V}}^{\dagger }(\xi ,g,u)\ , \label{nl2}
\end{eqnarray}
with
\begin{eqnarray}
u_{V}\in U_{V}(1)\ , & \quad &g\in SU_{c}(3)\ , \nonumber  \\h(\xi
,g,u)\in SU_{c}(2)\ , & \quad & h_{\widetilde{V}}(\xi ,g,u) \in
\widetilde{U}_{V}(1)\ .
\end{eqnarray}
It is convenient to define the following differential form:
\begin{eqnarray}
\omega _{\mu }=i{V}^{\dagger }D_{\mu }{V}\quad {\rm with}\quad
D_{\mu }{V}=(\partial _{\mu }-ig_{s}G_{\mu }){V}\ ,
\end{eqnarray}
with $G_{\mu }=G_{\mu }^{m}T^{m}$ the gluon fields while $g_{s}$
is the strong coupling constant. $\omega $ transforms according
to:
\begin{eqnarray}
\omega _{\mu }&\rightarrow & h(\xi ,g,u)\omega _{\mu }h^{\dagger
}(\xi ,g,u)+i\,h(\xi ,g,u)\partial {_{\mu }}h^{\dagger }(\xi ,g,u)
\nonumber \\ &+& i\,h_{\widetilde{V}}(\xi ,g,u)\partial _{\mu
}h_{\widetilde{V}}^{\dagger }(\xi ,g,u) \ . \nonumber
\end{eqnarray}
We decompose $\omega _{\mu }$ into
\begin{eqnarray}
\omega _{\mu }^{\parallel }=2S^{a}{\rm Tr}\left[ S^{a}\omega _{\mu
}\right] \quad {\rm and}\quad \omega _{\mu }^{\perp }=2X^{i}{\rm
Tr}\left[ X^{i}\omega _{\mu }\right] \ ,
\end{eqnarray}
$S^{a}$ are the unbroken generators of $H$, while
$S^{1,2,3}=T^{1,2,3}$ and $S^{4}=\widetilde{B}\,/\sqrt{2}$.

The most generic two derivative kinetic Lagrangian for the
goldstone bosons is:
\begin{eqnarray}
L=f^{2}a_{1}{\rm Tr}\left[ \,\omega _{\mu }^{\perp }\omega ^{\mu
\perp }\, \right] +f^{2}a_{2}{\rm Tr}\left[ \,\omega _{\mu
}^{\perp }\,\right] {\rm Tr} \left[ \,\omega ^{\mu \perp
}\,\right] \ . \label{dt}
\end{eqnarray}
The double trace term is due to the absence of the condition for
the vanishing of the trace for the broken generator $X^{5}$. It
emerges naturally in the non linear realization framework at the
same order in derivative expansion with respect to the single
trace term. In the unitary gauge these two terms correspond to the
five gluon masses \cite{CDS}.
\subsection{... and don't relax yet !}
 \label{fermions} For the fermions it is convenient to define the
dressed fermion fields
\begin{eqnarray}
\widetilde{\psi}={V}^{\dagger }\psi \ ,  \label{mq}
\end{eqnarray}
transforming as $\widetilde{\psi}\rightarrow
h_{\widetilde{V}}(\xi,g,u)h(\xi ,g,u)\,\widetilde{\psi}$. $\psi$
has the ordinary quark transformations (i.e. is a Dirac spinor).
\begin{figure}[ht]
\vskip -1.5 cm
\begin{center}
\includegraphics[angle=0, height=3cm, width =3.in]{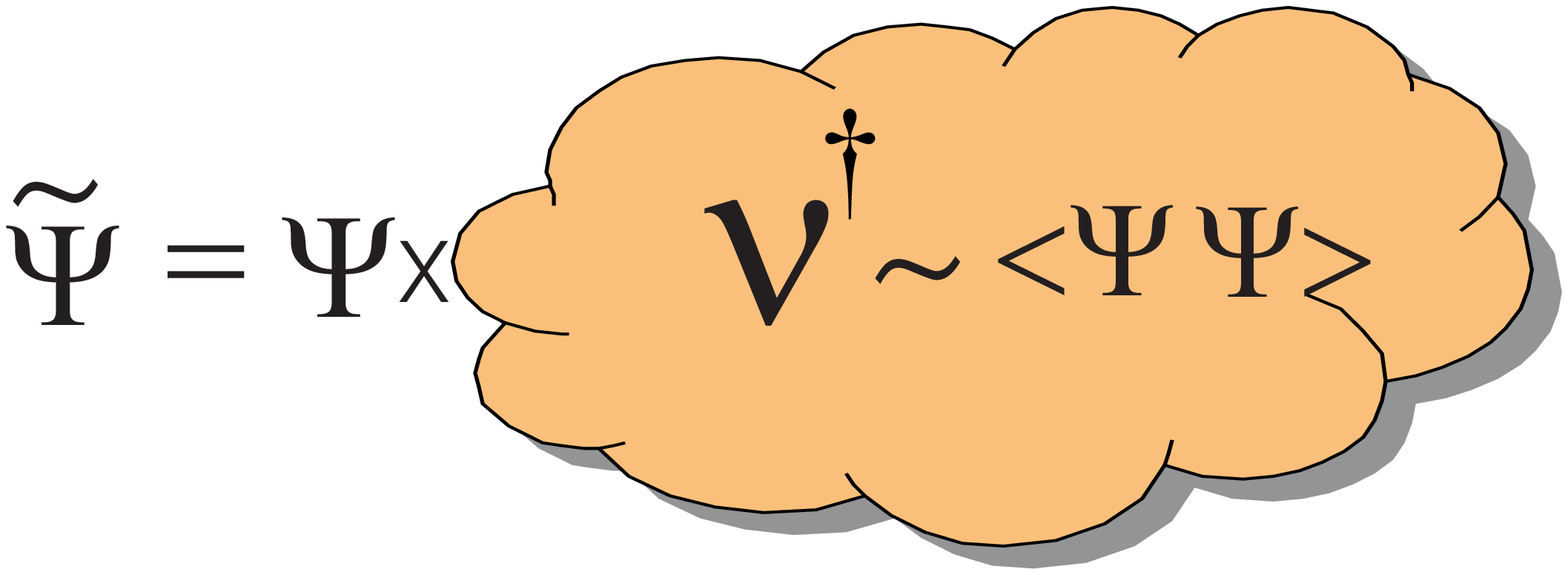}
\end{center}
\vskip -.6cm
\end {figure}
 Pictorially $\widetilde{\psi}$ can be viewed as a constituent
type field or alternatively as the bare quark field $\psi$
immersed in the diquark cloud represented by ${V}$. The non
linearly realized effective Lagrangian describing in medium
fermions, gluons and their self interactions, up to two
derivatives is:
\begin{eqnarray}
{L}&=&f^{2}a_{1}{\rm Tr}\left[ \,\omega _{\mu }^{\perp }\omega
^{\mu \perp }\,\right] +f^{2}a_{2}{\rm Tr}\left[ \,\omega _{\mu
}^{\perp }\,\right] {\rm Tr}\left[ \,\omega ^{\mu \perp }\,\right]
\nonumber \\ &+&b_{1}\overline{\widetilde{\psi }}i\gamma ^{\mu
}(\partial _{\mu }-i\omega _{\mu }^{\parallel })\widetilde{\psi
}+b_{2}\overline{\widetilde{\psi }} \gamma ^{\mu
}\omega _{\mu }^{\perp }\widetilde{\psi }  \nonumber \\
&+&m_{M}\overline{\widetilde{\psi }^{C}}_{i}\gamma
^{5}(iT^{2})\widetilde{\psi }_{j}\varepsilon ^{ij}+{\rm h.c.}\ ,
\end{eqnarray}
where $\widetilde{\psi }^{C}=i\gamma ^{2}\widetilde{\psi }^{\ast
}$, $i,j=1,2 $ are flavor indices and
\begin{eqnarray}
T^{2}=S^{2}=\frac{1}{2}\left(
\begin{array}{ll}
\sigma ^{2} & 0 \\
0 & 0
\end{array}
\right) \ .
\end{eqnarray}
Here $a_{1},~a_{2},~b_{1}$ and $b_{2}$ are real coefficients while
$m_{M}$ is complex.
{}From the last two terms, representing a Majorana mass term for
the quarks, we see that the massless degrees of freedom are the
$\psi _{a=3,i}$. The latter possesses the correct quantum numbers
to match the 't~Hooft anomaly conditions \cite{S}.

\section{The $SU_c(2)$ Glueball Lagrangian}
The $SU_c(2)$ gauge symmetry does not break spontaneously and
confines. Calling $H$ a mass dimension four composite field
describing the scalar glueball we can construct the following
lagrangian \cite{OS2}:
\begin{eqnarray}
S_{G-ball}&=&\int
d^4x\left\{\frac{c}{2}\sqrt{b}\,H^{-\frac{3}{2}}\left[\partial^{0}
H
\partial^{0}H - v^2
\partial^iH
\partial^iH\right]  \right. \nonumber \\ &-& \left. \frac{b}{2}
H\log\left[\frac{H}{\hat{\Lambda}^4}\right] \right\} \ .
\label{G-ball}
\end{eqnarray}
This Lagrangian correctly encodes the underlying $SU_c(2)$ trace
anomaly. The glueballs move with the same velocity $v$ as the
underlying gluons in the 2SC color superconductor. $\hat{\Lambda}$
is related to the intrinsic scale associated with the $SU_c(2)$
theory and can be less than or of the order of few MeVs
\cite{Rischke:2000cn} \footnote{According to the present
normalization of the glueball field $\hat{\Lambda}^4$ is
$v\,\Lambda^4$ with $\Lambda$ the intrinsic scale of $SU_c(2)$
after the coordinates have been appropriately rescaled
\cite{Rischke:2000cn,{OS2}} to eliminate the $v$ dependence from
the action \cite{Schafer}.} Once created, the light $SU_c(2)$
glueballs are stable against strong interactions but not with
respect to electromagnetic processes \cite{OS2}. Indeed, the
glueballs couple to two photons via virtual quark loops.
\begin{eqnarray}
 \Gamma\left[h\rightarrow
\gamma\gamma\right] \approx 1.2\times 10^{-2}
\left[\frac{M_h}{1~{\rm MeV}}\right]^5~{\rm eV} \ ,
\end{eqnarray}
where $\alpha=e^2/4\pi \simeq 1/137$. {}For illustration purposes
we consider a glueball mass of the order of $1$~MeV which leads to
a decay time $\tau\sim~5.5\times~10^{-14}s$. This completes the
effective Lagrangian for the 2SC state which corresponds to the
Wigner-Weyl phase.

Using this Lagrangian one can estimate the $SU_c(2)$ glueball
melting temperature to be \cite{Sannino:2002re}:
\begin{eqnarray}
T_c \leq \sqrt[4]{\frac{90 {v}^3}{2\,e\, \pi^2}}\, {\hat{\Lambda}}
< T_{CSC} \ .
\end{eqnarray}
Where $T_{CSC}$ is the color superconductive transition
temperature.
\begin{figure}[ht]
\begin{center}
\includegraphics[angle=0,height = 2.5cm, width = 2in]{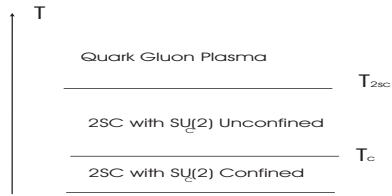}
\end {center}\caption[]{A zoom of the 2SC phases as function of temperature for fixed quark chemical potential.}
\end{figure}
The deconfining/confining $SU_c(2)$ phase transition within the
color superconductive phase is second order.
\section{Conclusions}
The Phase Diagram in Fig.~{\ref{tmu}} is just an educated guess of
the true QCD phase diagram. Indeed other interesting phases may
emerge. {}For example in the CFL phase the $K^{+}$ and $K^0$ modes
may be unstable for large values of the strange quark mass
signaling the formation of a kaon condensate
\cite{Schafer:2000ew}. Vortex solutions in dense quark matter due
to kaon condensation have been explored in \cite{Kaplan:uv}.

Another interesting avenue is the possibility of higher spin
condensates which can enrich the phase diagram structure of QCD
and QCD-like theories
\cite{{Sannino:2002wp},Sannino:2001fd,{Lenaghan:2001sd}}. Recent
lattice simulations seem to support these predictions
\cite{Alles:2002st,{Muroya:2002ry}} for 2 color QCD. If these
results are confirmed then for the first time we observe
spontaneous rotational breaking solely due to strongly interacting
matter.

An interesting part of the QCD phase diagram \cite{Halasz:1998qr}
which will be covered elsewhere is the temperature driven
confining-deconfining phase transition. Thanks to lattice
simulations we have a great deal of information
\cite{Boyd:1996bx}. Only very recently new methods have been
proposed which might help studying the phase diagram at non zero
chemical potential via lattice simulations \cite{Fodor:2002hs}.
However still much is left to be understood about the nature of
the transition of hot hadronic matter to a plasma of deconfined
quarks and gluons \cite{Satz:2002mg}. New effective Lagrangians
for the Polyakov loops \cite{Pisarski:2001pe} and the Polyakov
loops together with the hadronic states for the pure Yang-Mills
theory \cite{Sannino:2002wb} lead to a deeper understanding of the
properties of the underlying, temperature driven, deconfinement
transition\cite{AKF}.

\section*{Acknowledgments}
It is a pleasure to thank P.H.~Damgaard for discussions and
K.~Tuominen for discussions and careful reading of the manuscript.
This work is supported by the Marie--Curie fellowship under
contract MCFI-2001-00181.

\end{document}